\documentclass[aps,pra,superscriptaddress,amsmath,amssymb,showpacs,twocolumn,notitlepage]{revtex4-1}
\usepackage{amsmath,amssymb,amsfonts,latexsym,color,epsfig,textcomp}
\usepackage{subfigure}
\usepackage[latin1]{inputenc}
\usepackage{dcolumn}
\usepackage{bm}
\usepackage{graphicx}
\usepackage[english]{babel}
\usepackage{booktabs}
\usepackage{multirow}
\usepackage[table,xcdraw]{xcolor}

\DeclareMathOperator{\Rea}{Re}

\def\bbm[#1]{\mbox{\boldmath $#1$}}

\begin{document}
\title{Strong thermal and electrostatic manipulation of the Casimir force \\ in graphene multilayers}
\author{Chahine Abbas}
\affiliation{Laboratoire Charles Coulomb (L2C), UMR 5221 CNRS-Universit\'{e} de Montpellier, F- 34095 Montpellier, France}
\author{Brahim Guizal}
\affiliation{Laboratoire Charles Coulomb (L2C), UMR 5221 CNRS-Universit\'{e} de Montpellier, F- 34095 Montpellier, France}
\author{Mauro Antezza}
\email{Correspondance to: mauro.antezza@umontpellier.fr}
\affiliation{Laboratoire Charles Coulomb (L2C), UMR 5221 CNRS-Universit\'{e} de Montpellier, F- 34095 Montpellier, France}
\affiliation{Institut Universitaire de France - 1 rue Descartes, F-75231 Paris, France}
\date{\today}
%
%
\begin{abstract}

{We show that graphene-dielectric multilayers give rise to an  unusual tunability of the Casimir-Lifshitz forces, and allow to easily realize completely different regimes within the same structure. Concerning thermal effects, graphene-dielectric multilayers take advantage from the anomalous features predicted for isolated suspended graphene sheets, even though they are considerably affected  by the presence of the dielectric substrate. They can also archive the anomalous non-monotonic thermal metallic behavior by increasing the graphene sheets density and their Fermi energy. In addition to a strong thermal modulation occurring at short separations, in a region where the force is orders of magnitude larger than the one occurring at large distances, the force can be also adjusted by varying the number of graphene layers as well as their Fermi energy levels, allowing for relevant force amplifications which can be tuned, very rapidly and in-situ, by simply applying an electric potential. Our predictions can be relevant for both Casimir experiments and micro/nano electromechanical systems and in new devices for technological applications.}

\end{abstract}

\maketitle
%

The Casimir-Lifshitz pressure (CLP) occurring between closely-spaced bodies is a mechanical manifestation of both quantum vacuum and thermal fluctuations of radiation and matter fields \cite{Casimir,Lif56,DLP61}. It is the object of large theoretical and experimental interests \cite{CasimirBook} for both its fundamental and applicative implications. In particular, on the applicative side, such force has a clear impact in micro/nano (electro)mechanical systems (MEMS/NEMS), where it plays a dominant role at small separations \cite{Chan2001}. 
For parallel planar structures separated by a distance $d$ the CLP can be expressed as \cite{DLP61}    
\begin{multline}\label{Eq:Casimir_real}
\textrm{P}=-\frac{\hbar}{2\pi^2}  \int_{0}^{\infty}\textrm{d}\omega \coth{\left(\frac{\hbar\omega}{2k_BT}\right)}\;\times\\
\Rea\left[ \int_0^{\infty}\textrm{d}QQq_z\sum_{p}\left[ \frac{e^{-2 i q_z d}}{ R_p^{(1)}R_p^{(2)}}-1 \right]^{-1} \right],
\end{multline}
where $p=\textrm{TE, TM}$ stands for the two light polarizations (Transverse Electric and Transverse Magnetic), $q_z  = \sqrt{(\omega/c)^2-Q^2}$  and $R_p^{(i)}(Q,\omega,T)$ are the 
z-component of the vacuum wavevector and the reflection coefficient of bodies $i=1,2$, respectively. The integral is over the parallel-plane wavevector component $Q$. 

Equation \eqref{Eq:Casimir_real} shows that the CLP can be tuned by modifying the bodies' reflection coefficients or by varying the temperature $T$. In practice, thermal manipulation has been always considered as non effective: at short separations  ($d\leq1\mu$m), where the CLP is stronger, thermal effects are very small compared to vacuum ($T=0$K) ones. Remarkably, a thermal metal anomaly (TMA) has been predicted: for metals at intermediate separations ($\simeq 1\mu$m), contrary to dielectrics, the CLP decreases with increasing temperature \cite{Sernelius2000}. Thermal effects dominate only at very large separations $d\gg\lambda_T=\hbar c/(k_B T)$  ($\approx7\mu$m at room temperature) where the force reduces to the Lifshitz limit (for metals $\textrm{P}_{\textrm{Lif}}=-k_BT\zeta(3)/(8\pi d^3)$ \cite{Lif56}) and is already extremely weak and very hard to measure \cite{Antezza04,Harber05,Antezza05,Obrecht07,Lamoreaux}. For this reason, almost all research efforts focused on changing the reflection coefficients by using more complex geometries (recently large interest has been devoted to gratings \cite{Chan08, Lambrecht08,Mohideen2002,Guerout13,Decca13,Chan13,Messina15}) and/or materials (like topological insulators \cite{Grushin11}, metamaterials \cite{Milonni08}, switchable mirrors \cite{Iannuzzi04}, and others \cite{RMP}).

\begin{figure}[!ht]
\includegraphics[trim = 2cm 18.5cm 1cm 2cm,clip,width=0.48\textwidth]{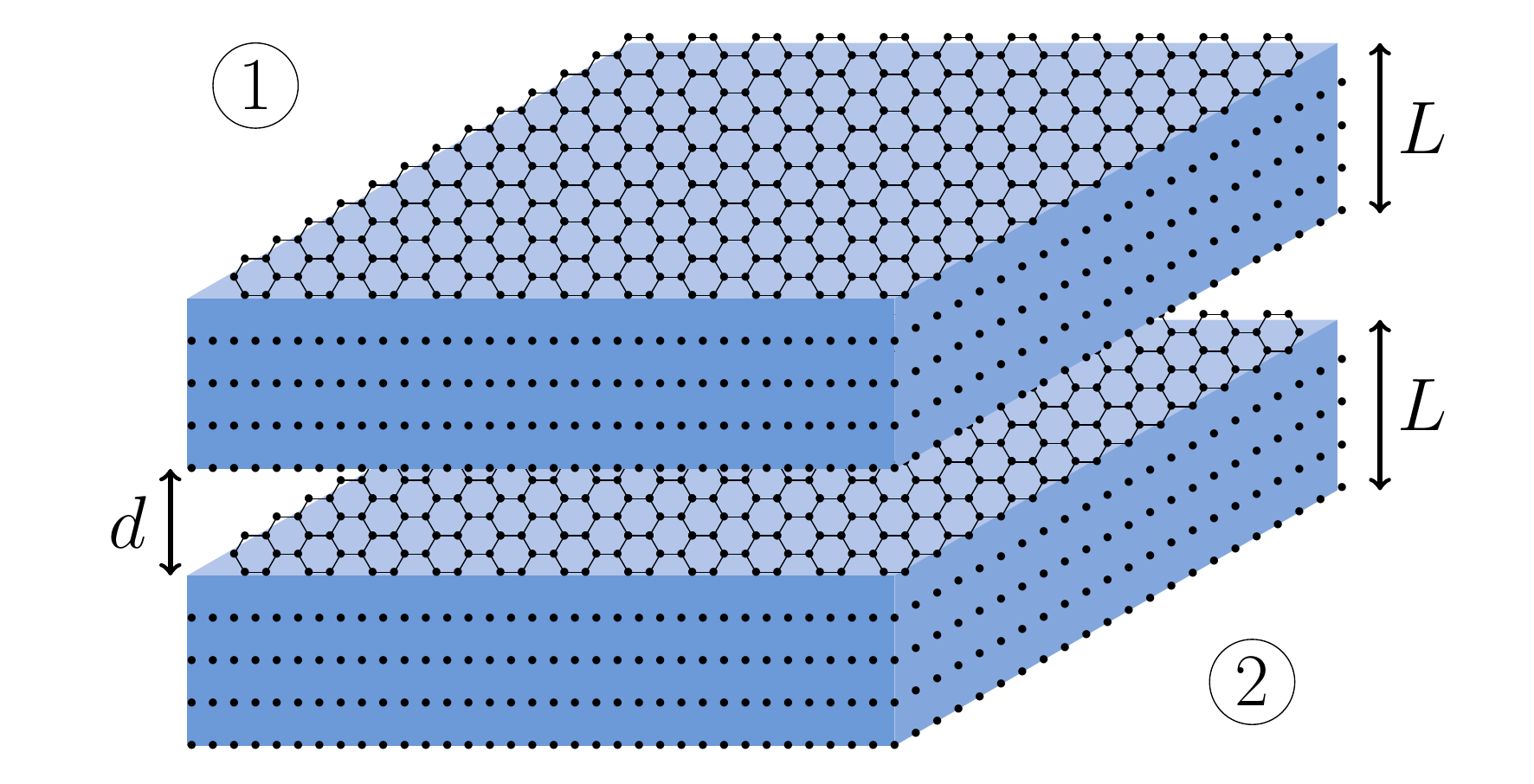}
\caption{\footnotesize (color online). Graphene-based multilayers scheme. \label{fig:figure_1}}
\end{figure}

Recently, the availability of graphene, with its peculiar transport and optical properties \cite{CastroRMP2008}, stimulated both theoretical \cite{Rubio06,Gomez09,Woods10,Svetovoy11,Mostepanenko14,Bordag15,Mostepanenko16} and experimental \cite{Banishev13} investigations of the CLP involving graphene sheets, with applications in nanophotonics and optomechanical systems \cite{Antezza2016}. 

\begin{figure}[!ht]
\includegraphics[width=0.5\textwidth]{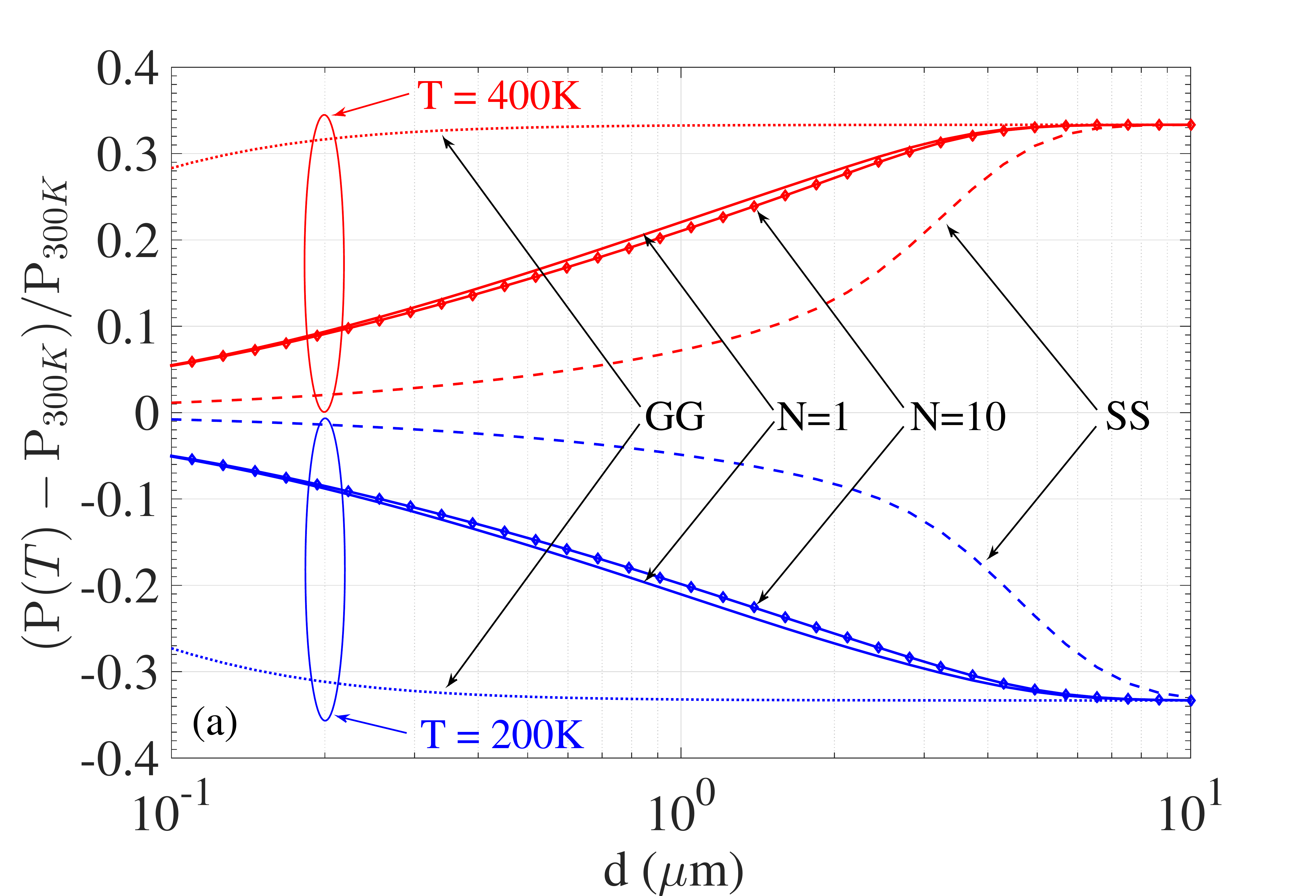}
\includegraphics[width=0.5\textwidth]{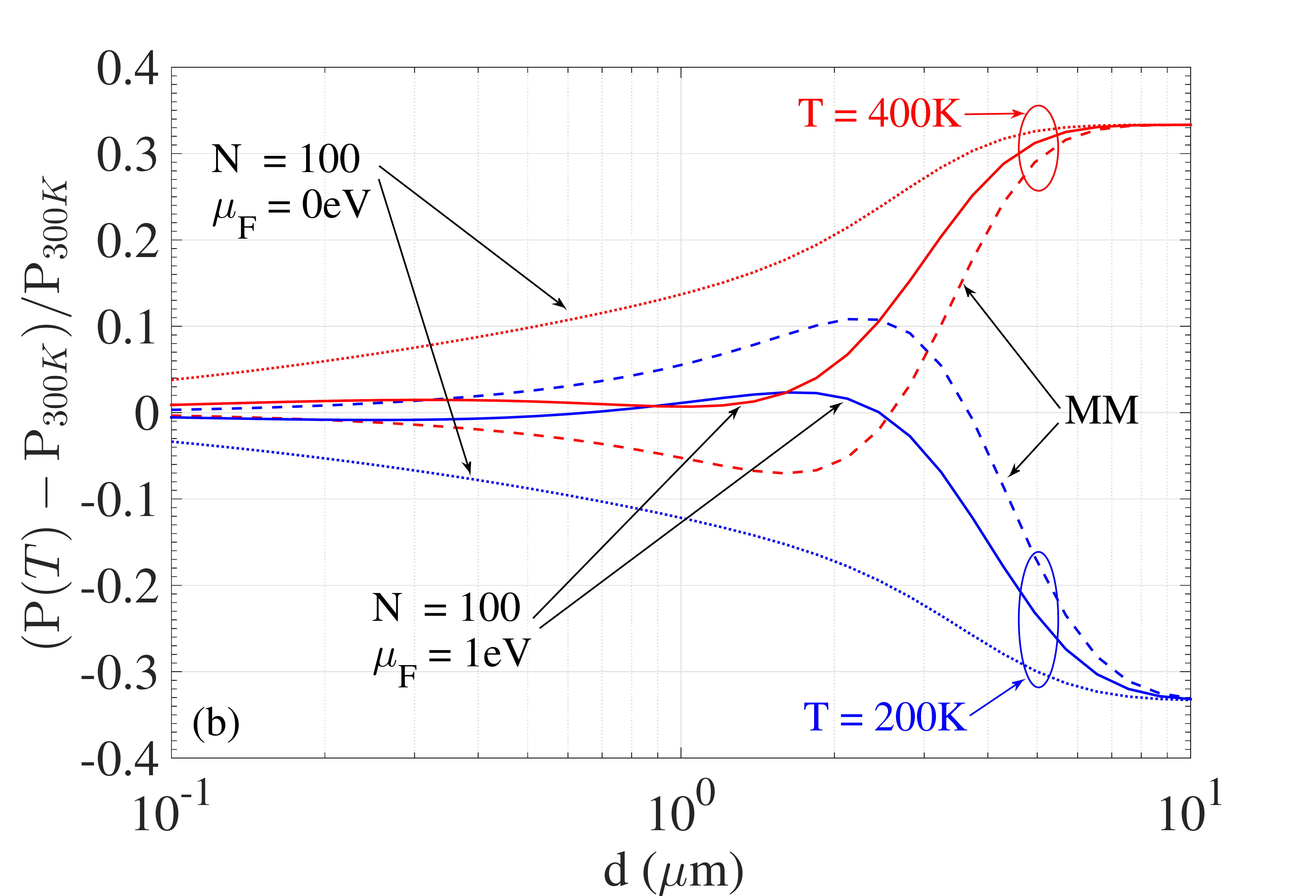}
\caption{\footnotesize {Relative variation of $\textrm{P}$ when $T$ varies [$T=200\,$K (blue lines) and $T=400\,$K (red lines)]. The CLP at $T=300K$ is taken as a reference, and $L=1\,\mu$m. (a): slabs (SS: dashed), suspended graphene sheets (GG: dotted), graphene-dielectric multilayers with $\mu_F=0\,$eV  ($N=1$:  solid, and $N=10$: solid with diamonds).  (b) gold (MM: dashed), graphene-dielectric multilayers with $N=100$ ($\mu=0\,$eV: dotted, and $\mu=1\,$eV: solid) \label{fig:figure_2}}.}
\end{figure}

{Remarkably, the CLP between two suspended parallel graphene sheets has been predicted \cite{Gomez09} to reach the Lifshitz metallic behavior $\textrm{P}_{\textrm{Lif}}$ at very small separations 
$d\gg \chi_T=\hbar v_F/(k_B T)\ll\lambda_T$, since the Fermi velocity is much smaller than $c$ ($v_F\simeq c/300$).  A natural question, then, is to which extent this striking thermal graphene anomaly (TGA) persists in typical realistic Casimir experimental conditions, which require the presence of substrates \cite{Banishev13,Lamoreaux} in a mixed graphene-dielectric configuration. This issue is also crucial for technological applications in MEMS/NEMS and in micro-optomechanical devices, calling for a  specific investigation due to the non-additive nature of the CLP. 

\begin{figure}[!ht]
\includegraphics[width=0.5\textwidth]{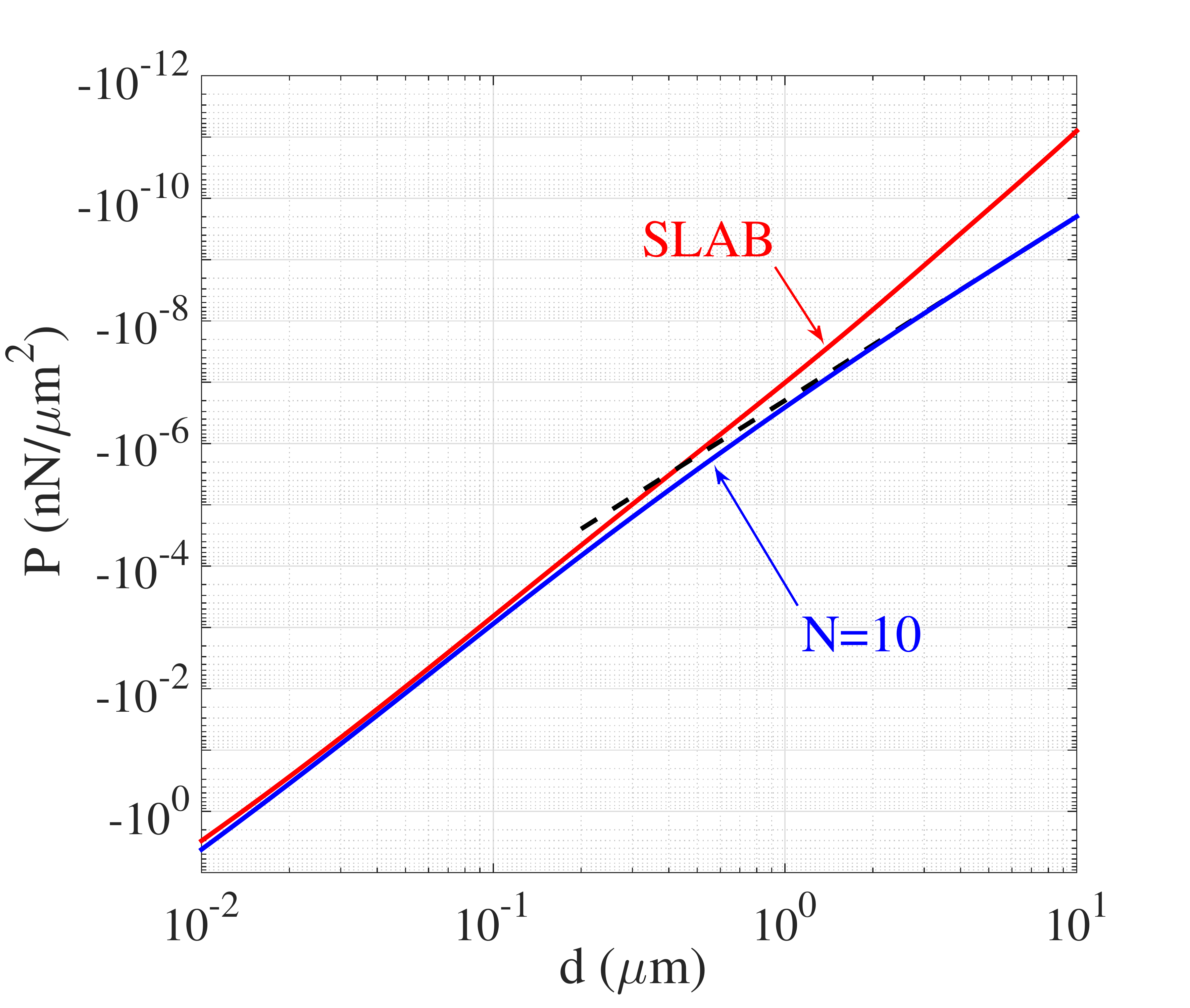}
\caption{\footnotesize CLP for $L=1\,\mu$m and $T=300\,$K, for different structures: SiO$_2$ slabs (red solid line),  graphene multilayers with $N=10$ and $\mu_F=0\,$eV (blue solid line), Lifshitz limit for metals $\textrm{P}_{\textrm{Lif}}$ (black dashed line). \label{fig:figure_3}}
\end{figure}

In this Letter, by simply introducing a dielectric substrate (we consider a general parallel-plane graphene-dielectric multilayer configuration), we propose a setting which allows several important modulations of the CLP and opens to genuine technological applications. Furthermore, it allows the compatibility with existing Casimir experiments and naturally guarantees the flatness and parallelism assumed in the model \cite{Gomez09}.

First, we show that the TGA strongly deviates from the ideal suspended-graphene configuration, still remaining large enough to thermally modulate the force at separations $\sim 200$nm, where the CLP is strong and typically measured. Second, we show that by increasing the density of the graphene layers in the dielectric host, we recover the TMA once the graphene is doped. Finally, we show that the same system allows an easy, strong and rapid CLP electrostatic tunability \emph{in-situ} by modulating the graphene conductivity with an applied voltage to the graphene sheets.

All these effects are particularly relevant for experiments since they allow, contrary to almost all known configurations, to dynamically change the force in the same experimental device without changing geometry or materials.}

\emph{Physical system and model -}  We consider the interaction between two identical parallel graphene multilayers imbedded in two dielectric slabs  separated by a distance $d$ (see figure \ref{fig:figure_1}). Each slab has a thickness $L$ and is loaded with $N_g$ equally spaced graphene sheets dividing the slab into $N=N_g-1$ layers. The dielectric layers are characterized by their permittivity $\varepsilon(\omega)$, and the graphene sheets by their conductivity $\sigma(\omega,T,\mu_F)$ (where $\mu_F$ is the Fermi level).  

\begin{figure}[!ht]
\includegraphics[width=0.48\textwidth]{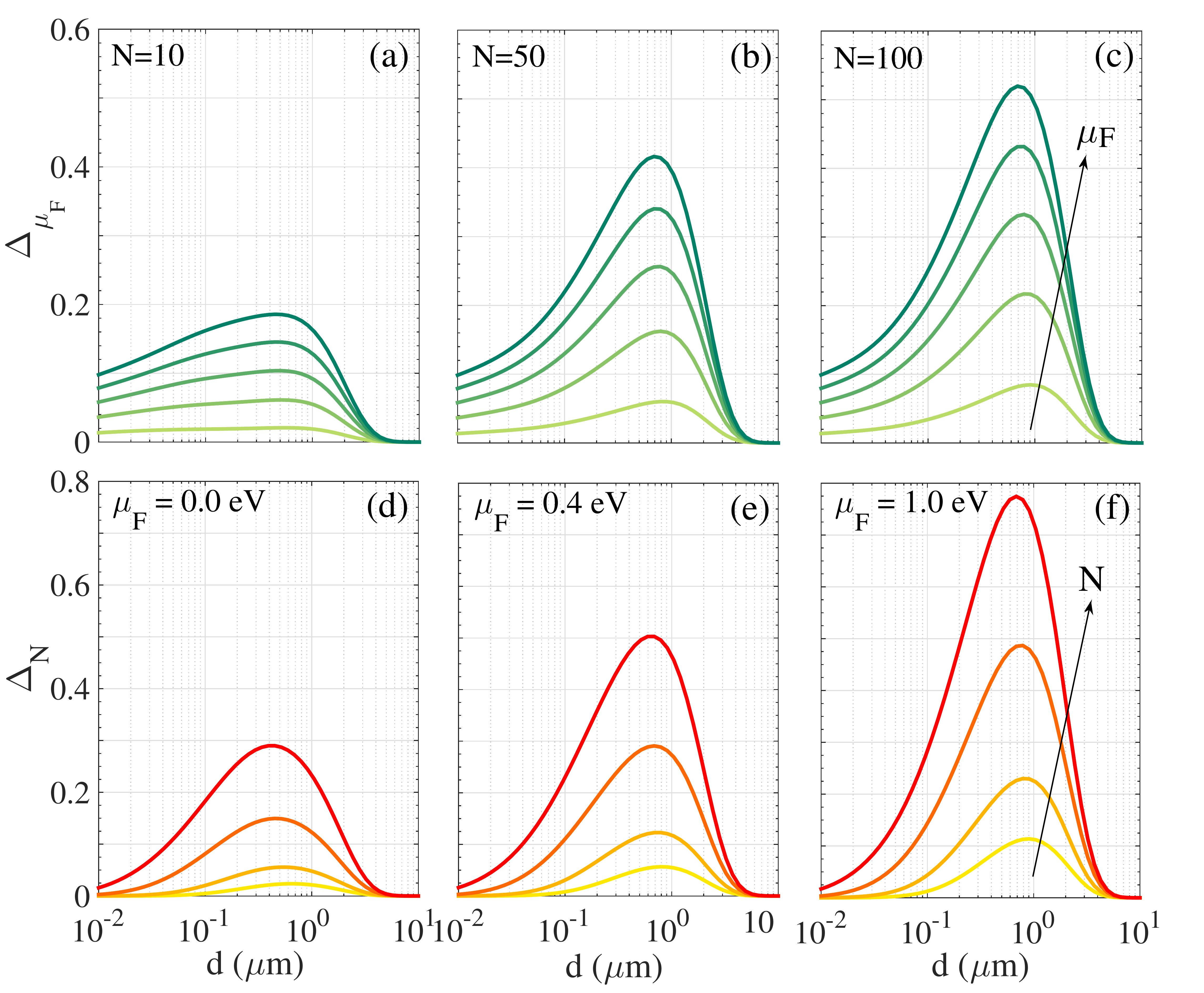}
\caption{\footnotesize Relative variation the CLP as a function of distance between the two multilayers, for $T=300\,$K and $L=1\,\mu$m. Panels (a)-(b)-(c): we plot $\Delta_{\mu_F}=\left| \left(P(\mu_F)-P(\mu_F=0)\right)/P(\mu_F=0) \right|$ for $\mu_F=0.2\,$eV, $0.4\,$eV, $0.6\,$eV, $0.8\,$eV and $1\,$eV, with $N=10$ (a), $N=50$ (b), $N=100$ (c). panels (d)-(e)-(f): we plot  $\Delta_{N}=\left| \left(P(N)-P(N=1)\right)/P(N=1) \right|$ for $N=10$, $20$, $50$ and $100$, with $\mu_F=0\,$eV (d), $\mu_F=0.4\,$eV (e) and $\mu_F=1\,$eV (f).\label{fig:figure_4}} 
\end{figure}

While Eq. \eqref{Eq:Casimir_real} is useful for understanding the roles of the different parameters, for computational efficiency we rather use its frequency complex-rotated version  ($\omega=i\xi_n$) \cite{DLP61}    
\begin{equation}\label{Eq:Casimir}
\textrm{P}=-\frac{k_BT}{\pi}  \sum_{n=0}^{\infty}\sideset{'}{}\int_0^{\infty}\textrm{d}QQq\sum_{p}\left[ \frac{e^{2 q d}}{ R_p^{(1)}R_p^{(2)}}-1 \right]^{-1},
\end{equation}
where the prime ${\prime}$ on the sum means that the $n=0$ term is divided by $2$, $\xi_n= \,2\pi n k_B T /\hbar$ are the Matsurbara frequencies, $q  = \sqrt{(\xi_n/c)^2+Q^2}$,  and $R_p^{(i)}(Q,i\xi_n,T)$ are the frequency-rotated reflection coefficients. In order to compute the graphene multilayers reflection coefficients  we implemented the scattering matrix algorithm (see \cite{SuppMat} for details) because of its outstanding stability with respect of all the parameters of the problem.

In the following we will consider SiO$_2$ slabs with permittivity $\varepsilon(\omega)=\varepsilon_\textrm{R}(\omega)+i \varepsilon_\textrm{I}(\omega)$ taken from \cite{Palik}, which at the Matsubara frequencies becomes  $ \varepsilon(i\xi_n)=1+\frac{2}{\pi}\int_0^\infty\frac{\omega\varepsilon_{\textrm{I}}(\omega)}{\omega^2+\xi_n^2}\textrm{d}\omega$ \cite{LLelec}.
The Graphene sheets conductivity $\sigma(\omega)=\sigma_\textrm{R}(\omega)+i\sigma_\textrm{I}(\omega)$ is the sum of the intra-band and inter-band contributions \cite{Falkovsky2008} (see also \cite{Falkovsky2007,Abajo2011,Ferrari2015}), and at Matsubara  frequencies takes the form {\cite{note_sigma}}:
\begin{eqnarray}
\sigma(i\xi_n)&=&\sigma_\textrm{intra}(i\xi_n)+\sigma_\textrm{inter}(i\xi_n), \label{eq:sigma}\\
\sigma_\textrm{intra}(i\xi_n)&=&\frac{8\sigma_0\;k_BT}{\pi(\hbar\xi_n+\hbar\Gamma)}\ln\left[2\;\cosh\left(\frac{\mu_F}{2k_BT}\right)\right], \nonumber\\
\sigma_\textrm{inter}(i\xi_n)&=&
\frac{\sigma_04\hbar\xi_n}{\pi}\int_0^{\infty}\frac{\mathcal{G}\left(x\right)}{(\hbar\xi_n)^2+4x^2}dx.\nonumber
\end{eqnarray}
Here, $\sigma_0=e^2/(4\hbar)$, $e$ is the electron charge, the Fermi level $\mu_F$ (typically between $0$ and $1\,$eV) can be modulated by applying a bias voltage or by chemical doping, $\mathcal{G}(x)=f(-x)-f(x)=\sinh(x/k_BT)/[\cosh(\mu_F/k_BT)+\cosh(x/k_BT)]$ with $f(x)=[(\exp[(x-\mu_F)/(k_BT)]+1]^{-1}$, and $\Gamma$ accounts for relaxation mechanisms  (we use $\Gamma=10^{13}$ rad/s).

\begin{figure}[!ht]
\includegraphics[width=0.47\textwidth]{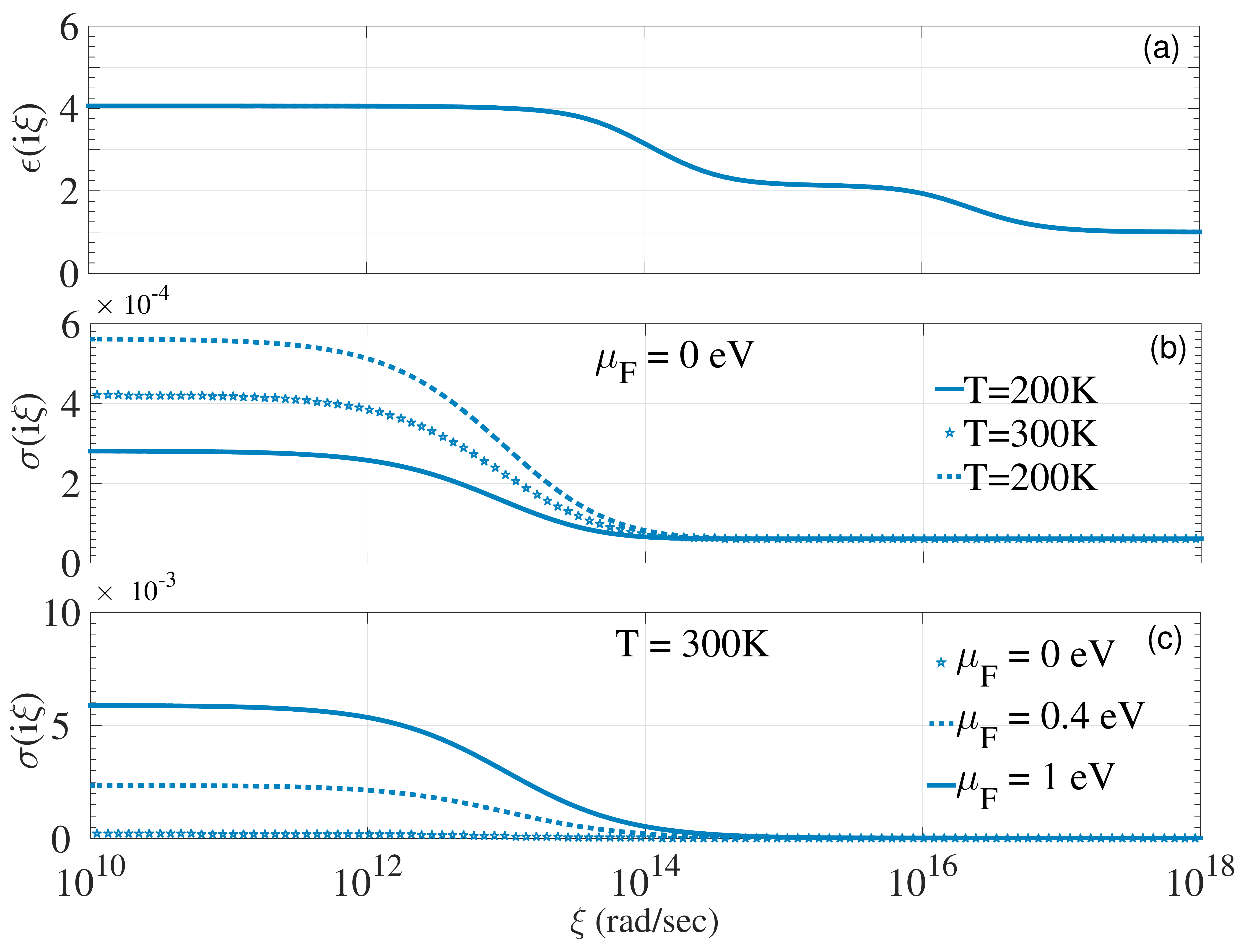}
\caption{\label{fig:figure_5} \footnotesize SiO$_2$ permittivity (a) and graphene conductivity (b) and (c) at imaginary frequencies.}
\end{figure}


\begin{figure}[!ht]
\includegraphics[width=0.47\textwidth]{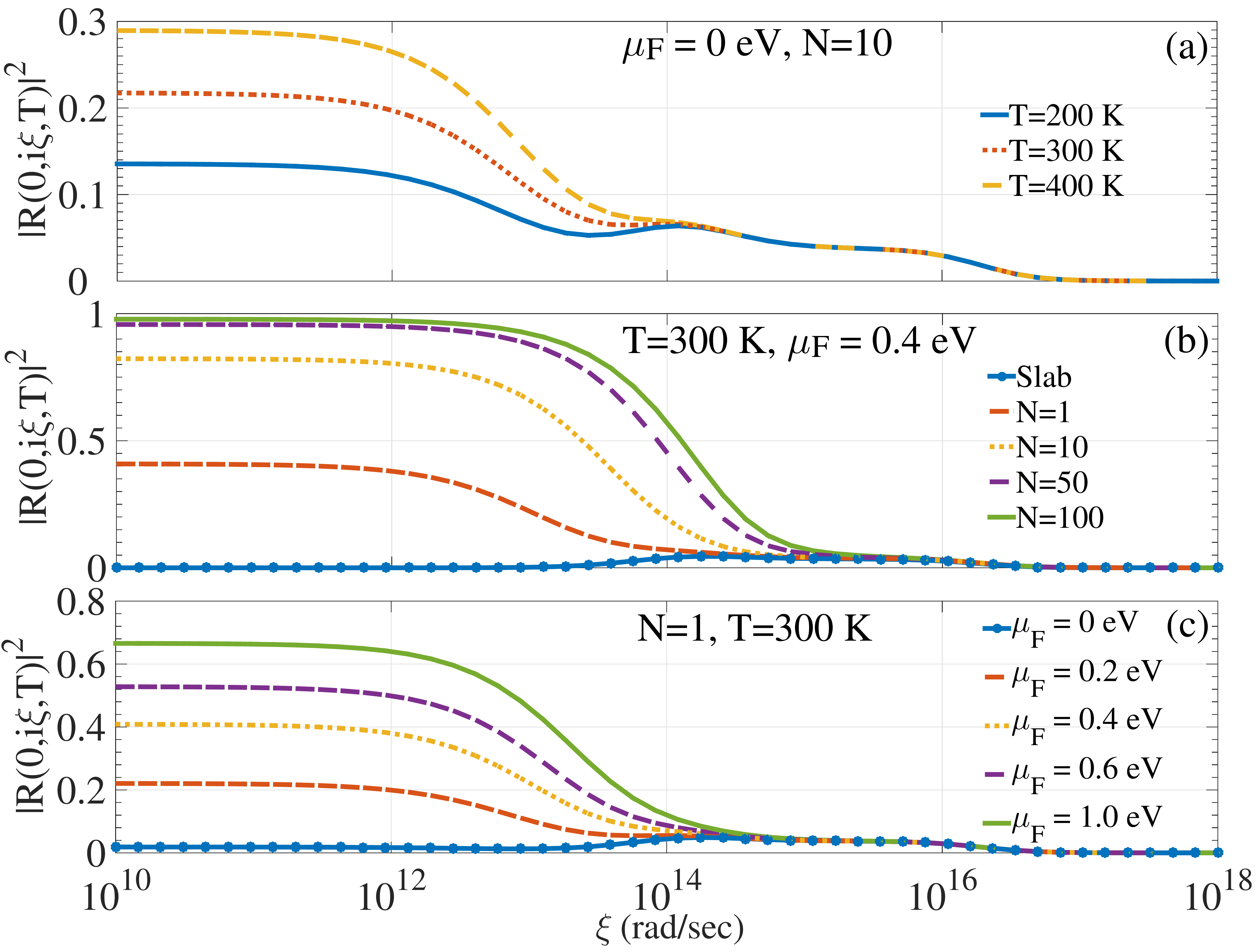}
\caption{\footnotesize TM reflectivity $|R|^2$ at normal incidence $(Q=0)$ at imaginary frequencies and for $L=1\,\mu$m. (a) $N=10$, $\mu_F=0\,$eV and $T$ varies (b) $T=300$K, $\mu_F=0.4\,$eV and $N$ varies (c) $T=300$K, $N=1$ and $\mu_F$ varies.\label{fig:figure_6}}
\end{figure}

\emph{Thermal and electrostatic modulation -} 
{We first focus on the influence of the temperature variation for graphene multilayer structures in figure \ref{fig:figure_2}, where we evaluate, as a function of the separation distance, the relative variation of the CLP for two different temperatures, namely $T=200$K and $T=400$K, with respect to the pressure at $T=300$K used as a reference. In panel (a) we compare the CLP between: two dielectric SiO$_2$ slabs, two parallel suspended graphene sheets, and two identical graphene-dielectric multilayers with $N=1$ (hence with $N_g=2$ graphene sheets each) and with $N=10$.}

{We see that the CLP in graphene-dielectric multilayers strongly deviates from that in suspended parallel graphene-graphene configuration, whose almost constant behavior in figure  \ref{fig:figure_2}(a) reflects the rapid TGA saturation of the CLP to the Lifshitz limit \cite{Gomez09}.
It is worth noticing that, in the case of slabs, to reach a $10\%$ relative variation, very large separations are required ($d\simeq2\mu$m both for $T=200\,$K and $T=400\,$K), where the total CLP is already negligible, making elusive the measurement of thermal effects. This appears clearly in figure \ref{fig:figure_3}, where at $d=2\mu$m the slab-slab CLP is $\simeq -10^{-8}$nN$/\mu$m$^2$. Remarkably, for graphene-dielectric multilayers (both for $N=1$ and $N=10$) a $10\%$ relative variation is already reached at $d\simeq200$nm. At this distance (which is typical in Casimir experiments) the CLP for graphene multilayers is $\simeq -10^{-4}$nN$/\mu$m$^2$  (see figure \ref{fig:figure_3}), which is four orders of magnitude larger than for the simple slabs configurations. This precisely opens to the possibility of measuring thermal effects, especially at small distances, and to thermally manipulate the force within standard Casimir experimental setups.  

In panel (b) we compare the CLP for: two gold slabs \cite{DrudeGold}, and two $N=100$ graphene-dielectric multilayers with $\mu=0\,$eV and $\mu=1\,$eV. We clearly see that for $\mu=0\,$eV the relative thermal variation for $N=100$ is weaker than for $N=1$ and $N=10$ (panel (a)), showing that by increasing $N$ the relative thermal effect decrease, while its absolute value increases  (see Fig. \ref{fig:figure_4}(d)), and that both $N=1$ and $N=10$ are almost equally good candidates to measure the CLP relative thermal variations. For $\mu=1\,$eV it becomes non-monotonic, acquiring the TMA behavior shown by gold. In that case the collective behavior of the 2D embedded graphene sheets makes the graphene-dielectric multilayer structure equivalent to an effective 3D metal.}

Let us now focus on other ways to tune the CLP which are offered by such structures. In the first line of figure \ref{fig:figure_4} we show how a change in the Fermi level $\mu_F$, which can be done in situ and dynamically, affects the CLP strength. We fix the number of layers (N=10, 50, 100 for panels (a), (b) and (c), respectively) and calculate, as a function of distance,  the relative variation of the CLP at increasing values of $\mu_F$, by normalizing with respect to the pressure at $\mu_F=0\,$eV. 

We see that, already with $N=10$, the relative variation can reach $20\%$ (panel(a)), and for $N=100$ a remarkable variation $>50\%$ can be obtained by continuously tuning $\mu_F$ up to $1\,$eV (panel (c)). In the second line of figure \ref{fig:figure_4} we show how much the CLP depends on N. We fix the Fermi level ($\mu_F=0, \,\,0.4, 1\,$ eV for panels (d), (e) and (f), respectively) and calculate the relative variation of the CLP at increasing values of $N$, normalizing with respect to the pressure with  $N=1$. We see that for $\mu_F=1\,$eV the relative variation for $N=100$ goes up to $\simeq80\%$  (panel (f)). It is worth stressing that in figure \ref{fig:figure_4}, by varying $N$ and/or $\mu_F$, the maximum variations are obtained at distances around $0.6\mu$m, {and become negligible at few microns, when the asymptotic universal regime $\textrm{P}_{\textrm{Lif}}$ is reached.}

In order to have more insight on the origin of the large CLP modulation (figures \ref{fig:figure_2} and \ref{fig:figure_3}) with respect to temperature $T$, Fermi level $\mu_F$ and the number of layers $N$, we first look, in figure \ref{fig:figure_5}, at the graphene conductivity $\sigma(i\xi)$ as a function of $T$ and $\mu_F$ . After, in figure \ref{fig:figure_6}, {we see how $\sigma(i\xi)$, jointly with the SiO$_2$ permittivity $\varepsilon(i\xi)$ and the variation of $N$, affect the multilayer reflection coefficient $R_p^{(i)}(Q,i\xi_n,T)$.}

The large thermal variation observed in figure \ref{fig:figure_2} derives from strong thermal variations of $\sigma(i\xi)$ (see Fig. \ref{fig:figure_5}(b)) which directly affect the multilayer TM reflectivity $|R|^2$ at normal incidence ($Q=0$) as shown in Fig. \ref{fig:figure_6}(a).
In Fig. \ref{fig:figure_5}(b)-(c) we see the interplay between $T$ and $\mu_F$ encoded in Eq. \eqref{eq:sigma}, which implies that a larger relative thermal variation is obtained for $\mu_F=0\,$eV {(for larger doping,  rapidly $\mu_F\gg k_BT\simeq10^{-2}$eV, implying no thermal conductivity effects). This explains the partial recovering of the TGA for $\mu_F=0\,$eV in Fig. \ref{fig:figure_2}(a-b), and, on the other side, explains that the TMA recovered in \ref{fig:figure_2}(b) for $\mu_F=1\,$eV is not due to thermal features of the Graphene sheets}. In Fig. \ref{fig:figure_6}(a) we see that the thermal variation of $\sigma(i\xi)$ affects the reflectivity mainly at frequencies smaller than $\simeq 10^{15}$rad/sec (which are the dominating frequencies in the Matsubara sum \eqref{Eq:Casimir}), while at larger frequencies the reflectivity is influenced only by the thermal-independent SiO$_2$ dielectric permittivity $\varepsilon(i\xi)$ given in Fig. \ref{fig:figure_5}(a).

{It is worth stressing that $|R(Q=0)|^2$ of Fig. \ref{fig:figure_6} is useful to understand the behavior of the CLP in general, where several Matsubara terms $\xi_n$ contribute to the sum \eqref{Eq:Casimir}. This is not the case for the large separations limit $d\rightarrow \infty$, for which one should consider \emph{only} the first Matsubara term $\xi_0=0$ rad/sec, and after perform the integration over $Q$. In that case, the reflectivities  reduce to the metallic limits $|R_{TM}(\xi_0)|^2=1$ and $|R_{TE}(\xi_0)|^2=0$ for any $Q\neq0$ (the $Q\rightarrow 0$ and $\omega\rightarrow 0$ limits ordering matters). In Fig. \ref{fig:figure_2}, the CLP for $N=1$ and $N=10$ at intermediate distances $d\simeq 1\mu$m  is not saturated by the single $\xi_0$ term (which would be enough for the suspended graphene configuration - dotted line) due to the mixed graphene-dielectric configuration.}

Let us now analyze the effect of varying both $\mu_F$ and $N$ on $|R|^2$: we see in Fig. \ref{fig:figure_6}(b) that adding and increasing the number of graphene sheets strongly modifies the reflectivity in a large range of frequencies $\lesssim 10^{15}$rad/sec, approaching more and more {an ideal metallic behavior} $|R|^2=1$ (while $|R|^2\simeq 0$ at small frequencies for simple slabs). Analogous variations of $|R|^2$ are shown if, at fixed values of $N$, the Fermi level increases, as shown in \ref{fig:figure_6}(c). The reflectivity increases considerably by increasing $N$ and/or $\mu_F$, which confers to graphene multilayer a {\emph{tunable metallic behavior}}, and explains the strong modulations of the CLP observed in  Fig. \ref{fig:figure_4}.

\emph{Conclusions-}
{We analyze, in terms of the graphene conductivity and of the structure reflection coefficients, both individual and collective effects of changing the temperature, the Fermi level and the number of graphene sheets on the CLP between graphene-dielectric multilayer structures. We exploit the fact that by changing $T$, $\mu$ and $N$ it is possible to modulate the graphene (semi)metallic features, and hence the reflectivity of the structure. For these structures we found that the CLP can strongly depend on temperature, implying a dramatic change with respect to both single suspended graphene sheets (more difficult to realize) and dielectric slabs, and allowing the measurement of thermal effects at small separations. Relevant similarities with normal 3D metals are found in some conditions.} We also show that a consistent modulation of the CLP can be obtained by varying the number of graphene sheets in the structure, or the Fermi level. This latter variation can be done by simply changing the electrostatic potential of the graphene sheets, and allows for a fast \emph{in-situ} tuning of the interaction, which is of clear experimental interest. A natural direct extension of this study is to consider non-ordered graphene-dielectric multilayer structures in order to further sculpt the CLP.  These findings offer several opportunities for both experimental Casimir investigations and for more applicative studies in micro/nano mechanical devices.

\emph{Acknowledgments-} We acknowledge Florian Bigourdan and {George Hanson} for useful discussions. 

\appendix

\section{Supplemental Material\\ The S-matrix algorithm for a multilayered structure with embedded graphene}\label{sec:SMatrix}
In order to calculate the multilayer reflection coefficient $R(Q,\omega)$ we use the so called $S$-matrix algorithm, well know for its effectiveness and stability. We present the algorithm for real frequencies $\omega$, but it remains valid also at imaginary Matsubara frequencies (needed in Eq.(2) of the main text) simply by setting $\omega=i\xi_n$.

\begin{figure}[htb] 
\includegraphics[trim = 0.5cm 13.75cm 1cm 2cm,clip,width=0.48\textwidth]{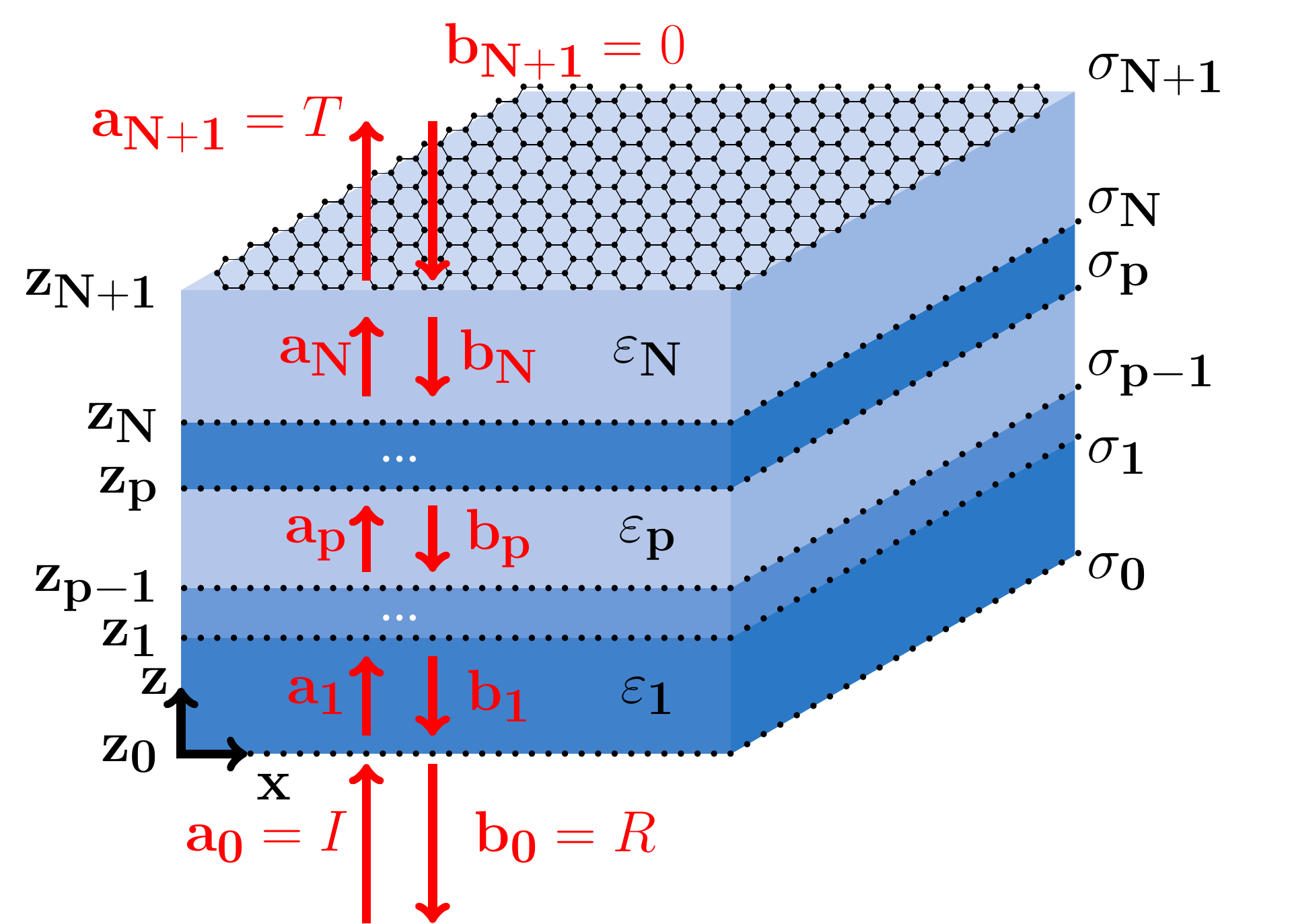}
\caption{\footnotesize (color online). Graphene-based multilayers scheme. 
\label{structure_2}}
\end{figure}

Let's consider the general multilayered structure, shown in figure \ref{structure_2}, made of $N$ dielectric layers and $N_g$ graphene sheets at their interfaces ($N_g=N+1$). Each layer is characterized by its width $h_p$, by its relative dielectric permittivity $\varepsilon_p$ and its relative magnetic permeability $\mu_p$ (in \cite{Chahine2017} we set $\mu_p=1$),  while each graphene sheet is characterized by its conductivity $\sigma_p$. We label each layer by its position number in the stack $p=1,...,N$ and label the lower and upper half spaces by $0$ and $N+1$ respectively. The whole structure is invariant in the $y$ direction and thus one can distinguish the two cases of polarization TE/TM according to this axis. Under the TE polarization, the electromagnetic field is such that ${\bf E}=(0,E_y,0)$ and ${\bf H}=(H_x,0,H_z)$ while for the TM case it is such that ${\bf E}=(E_x,0,E_z)$ and ${\bf H}=(0,H_y,0)$. Thus, for each polarization, the fields can be expressed through their non null $y$ component only; the other components being deduced from this latter through Maxwell's equations. 
With these notations, we express the fields in the $p^{th}$ medium in terms of plane waves solutions:
\begin{equation} \label{Eq:Fields}
 U_p(x,z)=e^{iQ x} \left(a_pe^{iq_p (z-z_{p-1})} + a_pe^{iq_p (z-z_{p-1})} \right)
\end{equation}
With $z_{-1}=0$ (by convention) and where $U_p(x,z)$ stands for $E_{py}(x,z)$ (respectively $H_{py}(x,z)$) in the TE (respectively TM) polarization case. Here $Q$ is the parallel component of the wave-vector and $q_p=\sqrt{k_0^2\varepsilon_p-Q^2}$ is its normal one, $k_0=\omega/c$ being the vacuum wave-number of the incoming plane wave. For a propagating incident wave, $Q$ can be related to the angle of incidence $\theta$ trough $Q=k_0\sqrt{\varepsilon_0} \sin{\theta}$.      
 
In order to compute the outgoing amplitudes $b_0=R $   and $a_{N+1}=T$ in terms of the incoming ones $a_0=I$  and $b_{N+1}=0$, we must take into account the boundary conditions at the different interfaces. These depend on the TE and TM polarization cases, and thus will be treated separately.

\subsubsection{TE polarization} 
For the $TE$ polarization case the boundary conditions can be expressed for each interface $z=z_p$ as follows:
\begin{equation} \label{Eq:TE_BondCond1}
\forall x \in \mathbb{R}: \left\{ \begin{array}{ll}
E_{py}(x,z_p)=E_{(p+1)y}(x,z_p) \\ H_{(p+1)x}(x,z_p)-H_{px}(x,z_p)=\sigma_pE_{py}(x,z_p).
\end{array} \right. 
\end{equation}
Then, using equation \eqref{Eq:Fields} and the Maxwell equation $H_x=(-i\omega\mu_0\mu_p)^{-1}\partial_zE_y$, we obtain:
 \begin{equation} \label{Eq:TE_BondCond2}
\left\{ \begin{array}{ll|}
\phi_pa_p+\phi_p^{-1}b_p=a_{p+1}+b_{p+1}  \\ q_{p+1}' \left( a_{p+1}-b_{p+1} \right) - q_{p}' \left( \phi_pa_p-\phi_p^{-1}b_p \right)=\\ \;\;\;\;\;\;\;\;\;\;\;\;\;\;\;\;\;\;\;\;\;\;\;\;\;\;\;\;\;\;\;\;\;\;\;\;-k_0 \eta_p \left(\phi_pa_p+\phi_p^{-1}b_p \right),
\end{array} \right. 
\end{equation}
where $q_p'=q_p/\mu_p$, $\phi_p=e^{iq_ph_p}$ ($h_p=z_{p+1}-z_p$ and $\phi_0=1$ by convention)  and $\eta_p=Z_0\sigma_p$, $Z_0$ being the electromagnetic impedance of vacuum. These boundary conditions constitute an algebraic set of $2N+2$ equations for the $2N+2$ unknowns $a_p, b_p$. One of the most efficient and stable ways to solve this latter system is to use the S-matrix algorithm. By definition, the S-matrix relates the outgoing amplitudes to the incoming ones:  

\begin{equation} \label{Eq:TE_BondCond_Smat1}
\left(\begin{array}{c}  b_p  \\  a_{p+1} \end{array} \right) = S_p^{\rm{TE}} \left(\begin{array}{c}  a_p  \\  b_{p+1} \end{array} \right).
\end{equation}

Therefore, we can deduce its expression easily from equations \eqref{Eq:TE_BondCond1}:

\begin{multline} 
\label{Eq:TE_BondCond_Smat2}
S_p^{\rm{TE}}=\dfrac{1}{q_{p+1}'+q_p'+k_0\eta_p} \;\times\\
\left(\begin{array}{cc} \phi_p^2(q_{p}'-q_{p+1}'-k_0\eta_p)  & 2\phi_p q_{p+1}' \\  2 \phi_p q_p'  & q_{p+1}'-q_p'-k_0\eta_p \end{array} \right).
\end{multline} 

Then chaining the successive S-matrices leads to the overall scattering matrix of the structure:
\begin{equation} \label{Eq:Chain} 
S^{\rm{TE}}=S_0^{\rm{TE}}\star...\star S_p^{\rm{TE}}\star...\star S_{N}^{\rm{TE}}
\end{equation}
where the $\star$ product $S=S^a\star S^b$ between two S-matrices $S^a$ and $S^b$ is
\begin{equation}  \label{Eq:ChainStar}
\left\{ \begin{array}{llll}
S_{11}=  S^a_{11}+S^a_{12}(1-S^b_{11}S^a_{22})^{-1}S^b_{11}S^a_{21}\\
S_{12}=  S^a_{12}(1-S^b_{11}S^a_{22})^{-1}S^b_{12}\\
S_{21}=  S^b_{21}(1-S^a_{22}S^b_{11})^{-1}S^a_{21}  \\
S_{22}=  S^b_{22}+S^b_{21}(1-S^a_{22}S^b_{11})^{-1}S^a_{22}S^b_{12}. \\
\end{array} \right.  
\end{equation}
\vspace{0.2cm}
Finally, the reflection and transmission coefficients are readily obtained from the global S-matrix:
\begin{equation} \label{Eq:Smat_coefs}
\left(\begin{array}{c}  R  \\ T \end{array} \right) = S^{\rm{TE}} \left(\begin{array}{c}  I  \\  0 \end{array} \right)
\end{equation}
so that the TE reflection coefficient we need for the Casimir-Lifshitz pressure calculation is simply the $(1,1)$ element of  $S^{\rm{TE}}$: $R_{\rm{TE}}^{(1)}(Q,\omega)=R_{\rm{TE}}^{(2)}(Q,\omega)=R/I=S_{11}^{\rm{TE}}$. {For completeness, the TE transmission coefficient will be $T_{\rm{TE}}^{(1)}(Q,\omega)=Te^{-iq_{N+1}L}/I=S_{21}^{\rm{TE}}e^{-iq_{N+1}L}$, where L is the size of the total multilayer structure, and where the phase factor $e^{-iq_{N+1}L}$ is introduced to have the transmission coefficient defined with respect to the $z_0$ plane, as for the reflection coefficient.} 
\subsubsection{$TM$ polarization} 
For the $\rm{TM}$ polarization case we follow the same procedure used for the TE case. We express the boundary conditions  for each interface $z=z_p$ as follows:
\begin{equation} \label{Eq:TM_BondCond1}
\forall x \in \mathbb{R}: \left\{ \begin{array}{ll}
E_{px}(x,z_p)=E_{(p+1)x}(x,z_p) \\ H_{(p+1)y}(x,z_p)-H_{py}(x,z_p)=-\sigma_pE_{px}(x,z_p).
\end{array} \right. 
\end{equation}
And now, by using equation \eqref{Eq:Fields} and the Maxwell equation $E_x=(i\omega\varepsilon_0\varepsilon)^{-1}\partial_zH_y$ we obtain:
\begin{equation} \label{Eq:TE_BondCond2}
\left\{ \begin{array}{ll}
q_p'( \phi_pa_p-\phi_p^{-1}b_p)=q_{p+1}'(a_{p+1}-b_{p+1})  \\ k_0 \left( a_{p+1}+b_{p+1} \right) - k_0 \left( \phi_pa_p+\phi_p^{-1}b_p \right)= \\ \;\;\;\;\;\;\;\;\;\;\;\;\;\;\;\;\;\;\;\;\;\;\;\;\;\;\;\;\;\;\;\;\;-\eta_p q_p'\left(\phi_pa_p-\phi_p^{-1}b_p \right),
\end{array} \right. 
\end{equation}
where $q_p'=q_p/\varepsilon_p$. The S-matrix can then be obtained:
\begin{widetext}
\begin{equation} \label{Eq:TE_BondCond_Smat2}
S_p^{\rm{TM}}=\dfrac{1}{k_0q_p'+k_0q_{p+1}'+\eta_pq_p'q_{p+1}'}\left(\begin{array}{cc}  \phi_p^2(k_0q_p'-k_0q_{p+1}'+\eta_pq_p'q_{p+1}' ) & 2k_0  \phi_p q_{p+1}' \\  2  k_0 \phi_p q_p'  & -k_0q_p'+k_0q_{p+1}'+\eta_pq_p'q_{p+1}' \end{array} \right).
\end{equation}
\end{widetext} 
The global TM S-matrix of the structure is then obtained by a chaining analogous to  Eq. \eqref{Eq:Chain}, and the TM reflection coefficient we need for the Casimir-Lifshitz pressure calculation is $R_{\rm{TM}}^{(1)}(Q,\omega)=R_{\rm{TM}}^{(2)}(Q,\omega)=R/I=S_{11}^{\rm{TM}}$. \\



\begin{thebibliography}{100}
\bibitem{DrudeGold} For gold, we used the Drude permittivity $\varepsilon(\omega)=1-\omega_p^2/(\omega^2+i\omega\gamma)$, with $\hbar\omega_p=9$eV and $\hbar\gamma=35$meV.
\end{thebibliography}
\end{document}